\documentclass[twocolumn,showpacs,preprintnumbers,prl,amsmath,amssymb]{revtex4}%PRL
\usepackage{graphicx}% Include figure files
\graphicspath{{../}{../Figs/}}
\usepackage{dcolumn}% Align table columns on decimal point
\usepackage{bm}% bold math
\usepackage{longtable}% tables which need more space than one page
\usepackage{array}
\usepackage{color}
\begin{document}

% \preprint{}
\title{Electrically tunable g-factors in quantum dot molecular spin states}
\author{M. F. Doty$^{1}$}
\email{doty@bloch.nrl.navy.mil}
\author{M. Scheibner$^{1}$}
\author{I. V. Ponomarev$^{1}$}
\author{E. A. Stinaff$^{1}$}
\author{A. S. Bracker$^{1}$}
\author{V. L. Korenev$^{2}$}
\author{T. L. Reinecke$^{1}$}
\author{D. Gammon$^{1}$}
\affiliation{$^{1}$Naval Research Laboratory, Washington, DC 20375,
USA} \affiliation{ $^{2}$A.F. Ioffe Physical Technical Institute,
St. Petersburg 194021, Russia}
\date{\today}% It is always \today, today,

\begin{abstract}
We present a magneto-photoluminescence study of individual
vertically stacked InAs/GaAs quantum dot pairs separated by thin
tunnel barriers. As an applied electric field tunes the relative
energies of the two dots, we observe a strong resonant increase or
decrease in the g-factors of different spin states that have
molecular wavefunctions distributed over both quantum dots. We
propose a phenomenological model for the change in g-factor based on
resonant changes in the amplitude of the wavefunction in the barrier
due to the formation of bonding and antibonding orbitals.
\end{abstract}

\pacs{75.40.Gb, 78.20.Ls, 78.47.+p, 78.67.Hc}% PACS, the Physics and Astronomy
\maketitle
% ***************************************************************
% *  Begin Main Text                                            *
% ***************************************************************
Quantum Dots and Quantum Dot Molecules (QDMs) have proven to be a
versatile medium for isolating and manipulating spins
\cite{KoppensScience05,PettaScience05}, which are of great interest
for quantum information processing \cite{LossPRA98,VrijenPRA00}. In
particular, photoluminescence (PL) spectra have been used in
self-assembled QDMs to observe coherent tunneling
\cite{StinaffScience06, KrennerPRL05,KrennerCondMat06,OrtnerPRL05}
and identify spin interactions through fine structure
\cite{Scheibner}. Electrical control of isolated spins through their
g-factors is highly desireable for implementation of quantum gate
operations. To date, electrical control of g-factors has only been
observed in ensembles of electrons in quantum wells by shifting the
electron wavefunctions into different materials \cite{SalisNature01,
PoggioPRB04, JiangPRB01, LinPhysicaE04}. In this Letter we present a
striking electric field resonance in the g-factor for molecular spin
states confined to a single quantum dot molecule.

\begin{figure}[htb]
\includegraphics[width=8.6cm]{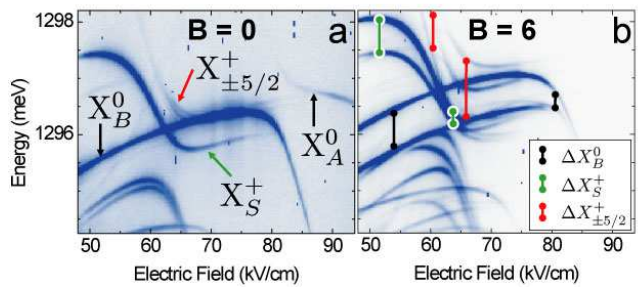}
\caption{(a) B = 0 T photoluminescence spectra of a single QDM. The
complex pattern of anticrossings arises from the formation of
molecular spin states. (b) At B = 6 T, the molecular spin states
have a Zeeman splitting (bars) that depends strongly on the applied
electric field (F).} \label{zeroandsix}
\end{figure}

To our knowledge this is the first observation of electrical control
over the g-factor for a single confined spin. Moreover, the
isolation of a single QDM allows us to spectrally resolve and
identify individual molecular spin states that have different
g-factor behaviors. In Fig.\ \ref{zeroandsix}a we indicate molecular
spin states of both the neutral exciton ($X^0$, one electron
recombining with one hole) and positive trion ($X^+$, electron-hole
recombination in the presence of an extra hole) at zero magnetic
field. The different electric field dependences of the g-factors for
these states is apparent in Fig.\ \ref{zeroandsix}b, where the
splitting of PL lines increases for some molecular spin states and
decreases for others. This electric field dependence is nearly an
order of magnitude larger than previously reported in quantum wells
\cite{SalisNature01, PoggioPRB04, JiangPRB01, LinPhysicaE04}. The
effect arises from the formation of bonding and antibonding
orbitals, which results in a change in the amplitude of the
wavefunction in the barrier at resonance.

Our QDMs consist of two vertically stacked self-assembled InAs dots
truncated at a thickness of 2.5 nm and separated by a 2 nm GaAs
tunneling barrier \cite{growthdetails}. As an applied electric field
tunes the relative energies of the two dots, strong tunnel coupling
between the hole states creates the molecular spin states. Unlike
samples with a thicker tunnel barrier \cite{StinaffScience06}, the
states retain molecular character throughout the observed range of
electric fields. We present data from a single molecule, but the
universality of the behavior has been verified by detailed studies
of 7 other molecules from the same sample. We first explain the
spectra and molecular spin states at B = 0 T. We then describe the
magnetic field dependence and propose a phenomenological model for
the electric field dependent Zeeman splitting.

\begin{figure}[htb]
\includegraphics[width=8.6cm]{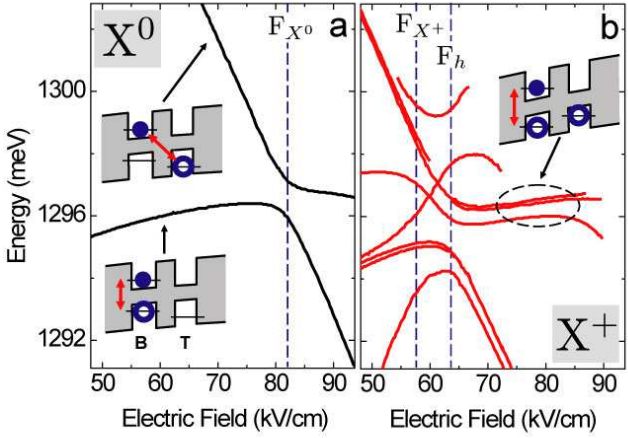}
\caption{Energies extracted from Fig.\ \ref{zeroandsix}a. (a) $X^0$
lines anticross at $F_{X^0}$, where the direct (lower inset) and
indirect (upper inset) transitions are degenerate. (b) $X^+$ initial
states anticross at $F_{X^+}$ and near 90 kV/cm. Final (hole) states
anticross at $F_h$. Inset: charge distribution for the circled
singlet and triplet transitions.} \label{Spectra}
\end{figure}

In Fig.\ \ref{Spectra} we show all PL lines from $X^0$ and $X^+$ at
B = 0 T. These lines are identified by their relative energies, the
power dependence of their intensities and the electric field
dependence of the anticrossings \cite{StinaffScience06}. The $X^0$
lines (Fig.\ \ref{Spectra}a) show a clear anticrossing at $F_{X^0}$.
The anticrossing arises from tunnel coupling between the ground
state hole levels in each dot, which forms bonding ($X^0_B$) and
antibonding ($X^0_A$) orbitals. The electron remains in the bottom
dot throughout the range of electric fields considered here
\cite{StinaffScience06}. The $X^+$ lines (Fig.\ \ref{Spectra}b) have
a more complicated pattern because anticrossings occur in both the
initial (one electron and two holes) and final (one hole) states
\cite{StinaffScience06, Scheibner}.

To explain the origin of the $X^+$ molecular spin states we turn to
the Hamiltonians. The basis states will be identified as
$^{e_B,e_T}_{h_B,h_T}X^Q_k$, where $e_B$ [$e_T$] are the spins of
electrons ($\pm1/2$: $\uparrow$ or $\downarrow$) and $h_B$ [$h_T$]
the spins of heavy holes ($\pm3/2$: $\Uparrow$ or $\Downarrow$) in
the bottom [top] dot. X indicates an exciton (h a single hole) and Q
is the net charge. k is the total spin projection. Singlets, which
have total spin $\pm1/2$, will be denoted $X_S$ to distinguish them
from the $\pm1/2$ triplets.

\begin{figure}[htb]
\includegraphics[width=8.6cm]{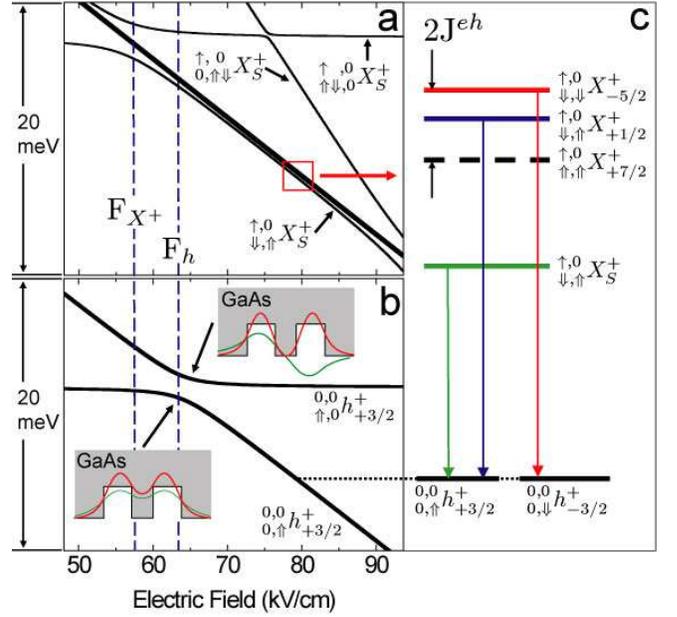}
\caption{Calculated zero magnetic field energies of spin-degenerate
(a) initial states ($X^+$) and (b) final states ($h^+$)
\cite{numericalvalues}. Labels indicate the dominant basis state for
easy comparison to Eq.\ \ref{X+Hamiltonian}. Insets in (b) show the
bonding (lower) and antibonding (upper) wavefunctions (green) and
probability distributions (red).  (c) Fine structure of singlet and
triplet states at F = 79.2 kV/cm. Arrows indicate optically allowed
transitions to the bonding final state.} \label{X+States}
\end{figure}

The final state has only a single hole with relative energies given
by the diagonalization of the Hamiltonian:
\begin{equation}
\label{hHamiltonian} \widehat{H}^h=\left(\begin{array}{cc}
  0 & t_h \\
  t_h & e \tilde{d} F \\
\end{array}\right)
\end{equation}
in the spin degenerate $^{0,0}_{\Uparrow,0}h^+_{+3/2}$,
$^{0,0}_{0,\Uparrow}h^+_{+3/2}$ basis, where $t_h$ is the tunneling
energy for a bare hole, $\tilde{d}$ is the distance between dot
centers and $F$ is the applied electric field. The energies of the
final states as a function of electric field are shown in Fig.\
\ref{X+States}b. The formation of bonding and antibonding states at
the anticrossing point is illustrated.

The initial state ($X^+$) contains one electron and two holes. For
simplicity we present only the electron-spin-up case, which is
degenerate with the spin-down case at zero magnetic field. If both
holes are in the same dot, the Pauli principle requires singlet
configurations: $^{\uparrow\hspace{5pt},0}_{\Uparrow
\Downarrow,0}X^+_S$ (both in the bottom dot) or
$^{\uparrow,\hspace{3pt}0}_{0,\Uparrow \Downarrow}X^+_S$ (both in
the top dot). If there is one hole in each dot, singlet
($^{\uparrow,0}_{\Downarrow, \Uparrow}X^+_S$) and triplet
($^{\uparrow,0}_{\Downarrow, \Uparrow}X^+_{+1/2}$,
$^{\uparrow,0}_{\Downarrow, \Downarrow}X^+_{-5/2}$,
$^{\uparrow,0}_{\Uparrow, \Uparrow}X^+_{+7/2}$) configurations are
possible. By $^{\uparrow,0}_{\Downarrow, \Uparrow}X^+_S$ we mean the
antisymmetric hole spin wavefunction ($\Downarrow,\Uparrow -
\Uparrow, \Downarrow$). $^{\uparrow,0}_{\Downarrow,
\Uparrow}X^+_{+1/2}$ means the symmetric hole spin wavefunction
($\Downarrow, \Uparrow + \Uparrow,\Downarrow$). The relative
energies of the initial states are given by the Hamiltonian:

\begin{equation}
\label{X+Hamiltonian}\widehat{H}^{X^+}=
\left(\begin{array}{ccc@{\hspace{-2pt}}c@{\hspace{-2pt}}c@{\hspace{-2pt}}c}
  \Gamma_1 & t_{X^+} & 0 & 0 & 0 & 0 \\
  t_{X^+} & e \tilde{d} F & J^{eh} & 0 & 0 & t_{X^+} \\
  0 & J^{eh} & E_{+1/2} & 0 & 0 & 0 \\
  0 & 0 & 0 &  E_{-5/2} & 0 & 0 \\
  0 & 0 & 0 & 0 & E_{+7/2}  & 0 \\
  0 & t_{X^+} & 0 & 0 & 0 & 2 e \tilde{d} F + \Gamma_2 \\
\end{array}\right)
\end{equation}
in the $^{\uparrow\hspace{5pt},0}_{\Uparrow \Downarrow,0}X^+_S$,
$^{\uparrow,0}_{\Downarrow, \Uparrow}X^+_S$,
$^{\uparrow,0}_{\Downarrow, \Uparrow}X^+_{+1/2}$,
$^{\uparrow,0}_{\Downarrow, \Downarrow}X^+_{-5/2}$,
$^{\uparrow,0}_{\Uparrow, \Uparrow}X^+_{+7/2}$,
$^{\uparrow,\hspace{3pt}0}_{0,\Uparrow \Downarrow}X^+_S$ basis.
$\Gamma_1$ and $\Gamma_2$ are due to Coulomb interactions, $t_{X^+}$
is the Coulomb-corrected tunneling energy for a hole in the presence
of an electron and additional hole, $E_k = e \tilde{d} F + mJ^{eh}$
with $m=0,+1,-1$ for $k=+1/2,-5/2,+7/2$. $J^{eh}$ is the exchange
energy between an electron and hole in the same dot
\cite{StinaffScience06, Jhh}. Diagonalizing Eq.\ \ref{X+Hamiltonian}
gives the energies of the initial states, which are plotted as
functions of electric field in Fig.\ \ref{X+States}a.

Because tunneling is a spin conserving process, only the
$^{\uparrow,0}_{\Downarrow, \Uparrow}X^+_S$ singlet state can tunnel
couple with $^{\uparrow\hspace{5pt},0}_{\Uparrow \Downarrow,0}X^+_S$
and $^{\uparrow,\hspace{3pt}0}_{0,\Uparrow \Downarrow}X^+_S$, which
must be singlets because the two holes are in the same dot. These 3
singlets are therefore strongly mixed to create molecular orbital
states that anticross near 57 kV/cm ($F_{X^+}$) and 90 kV/cm. Unlike
the singlet states, the triplet states do not mix and are not
affected by these anticrossings \cite{STmixing}. This creates a
``kinetic" splitting between triplet and singlet states
\cite{Fazekas99}. An example is indicated by the dashed oval in
Fig.\ \ref{Spectra}b, where the lower energy singlet line remains
separated from the two optically allowed triplet lines, which are
split by electron-hole exchange. The fine structure of the
corresponding states is shown in Fig.\ \ref{X+States}c.

\begin{figure}[tb]
\includegraphics[width=8.6cm]{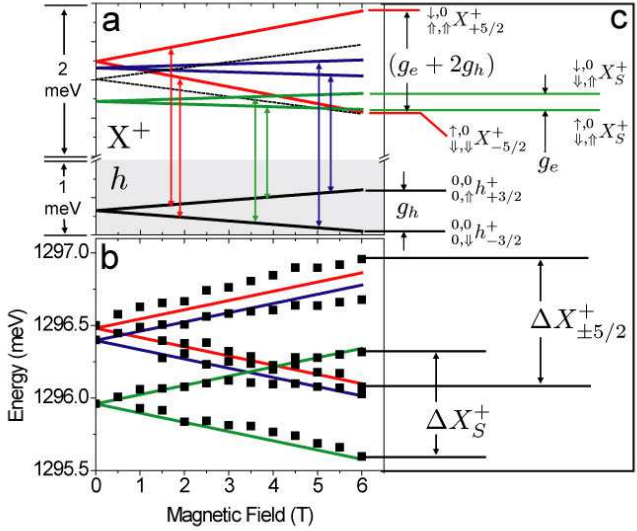}
\caption{(a) Magnetic field dependence of initial ($X^+$) and final
($h$) states for the singlet and triplet transitions at an electric
field of 79.2 kV/cm (schematic Fig.\ \ref{X+States}c). Vertical
lines indicate spin allowed recombinations. (b) Calculated (colored
lines) and experimentally observed (black points) PL energies. (c)
Zeeman splitting of PL lines ($\Delta$) and initial and final states
for $X^+_{\pm5/2}$ (red lines) and $X^+_S$ (green lines).}
\label{BStates}
\end{figure}

By adding the Zeeman interaction to the Hamiltonian
\cite{BayerPRB00}, we calculate the magnetic field dependence of the
molecular spin states. Fig.\ \ref{BStates}a shows the states of
Fig.\ \ref{X+States}c, which split into doublets with an applied
longitudinal magnetic field. The final states are simply the two
spin orientations of a single hole, split by $g_h$. The splitting of
initial states depends on their spin configuration. Due to the two
parallel hole spins, $X^+_{\pm5/2}$ has a large splitting given by
$g_e+2g_h$ while $X^+_{\pm7/2}$ is split by $-g_e+2g_h$. In
contrast, the $\pm1/2$ singlet ($X^+_S$) and triplet
($X^+_{\pm1/2}$) have oppositely paired hole spins and therefore a
small splitting given by $g_e$. The g-factor for PL transitions is
given by the difference in g-factor between the initial and final
states. Away from the electric field resonances, the g-factor is
$g_T = g_e+g_h$, as indicated by the vertical lines.

Using the model described below, we obtain $g_T=-2.2$. To plot the
initial and final states in Fig.\ \ref{BStates}a we have taken the
relative weights of the electron and hole g-factors to match those
obtained by Bayer: $g_e=-0.6$ and $g_h=-1.6$ \cite{BayerPRB00}. The
calculated energies of the transitions are shown by the lines in
Fig.\ \ref{BStates}b. The experimentally observed PL energies are
given by the symbols, with the diamagnetic shift (10.9
$\mu$eV/T$^2$) subtracted.

\begin{figure}[tb]
\includegraphics[width=8.6cm]{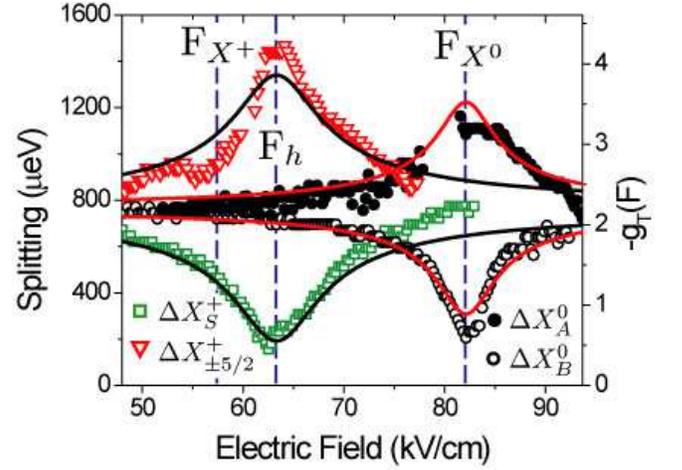}
\caption{Zeeman splitting and corresponding g-factor, $g_T$, as a
function of electric field at B = 6 T.} \label{gfactor}
\end{figure}

The g-factor resonances are clearly evident in Fig.\ \ref{gfactor},
where the symbols plot the measured energy splitting of the $X^0$
and two $X^+$ Zeeman doublets at B = 6 T as a function of electric
field. Strong enhancement or suppression of the splitting is
observed at $F_h$ and $F_{X^0}$. All of the data can be
qualitatively explained by a phenomenological model of the formation
of bonding and antibonding orbitals, which results in resonant
changes in the amplitude of the wavefunction in the barrier. We
focus first on transitions involving a bonding orbital (open
symbols), which have an increased amplitude of the wavefunction in
the barrier (Fig.\ \ref{X+States}b lower inset).

The wavefunction for the bonding orbital of a single hole can be
written as $\Psi_B = a|1\rangle+b|2\rangle$, where $|1\rangle$ and
$|2\rangle$ are the basis states of Eq.\ \ref{hHamiltonian}, the
wavefunctions for holes localized in the two different dots. The
coefficients $a$ and $b$ are functions of electric field determined
by Eq.\ \ref{hHamiltonian}. The electric field-dependent g-factor
for a hole in a bonding orbital is given by
$g^B_h(F)=\langle\Psi_B|g_h(z)|\Psi_B\rangle$, where $g_h(z)$ is the
hole g-factor as a function of position in the sample. $g_h(z)$ is
taken as a phenomenological parameter, in part because the degree of
alloying between the nominally pure InAs dots and GaAs barrier is
unknown \cite{gfactors}.

If we assume that the g-factors for the holes localized in each dot
are the same, we get $g^B_h(F)=g_h+2abg_{12}$, where
$g_{12}=\langle1|g_h(z)|2\rangle$ gives the contribution from the
amplitude of the wavefunction in the barrier. In the case of the
$X^+_S$ singlet (shown in Figs. \ref{X+States}c and \ref{BStates}c),
the initial states are split by $g_e$, so the total transition
g-factor is given by
\begin{equation}\label{gfactormodel}
g^B_T(F)
=g_e+g^B_h(F)=g_T+\frac{2t_hg_{12}}{\sqrt{e^2\tilde{d}^2(F-F_h)^2+4t_h^2}}
\end{equation}
where the explicit form for $2ab$ determines the lineshape centered
around the anticrossing point $F_h$.

The lower black line in Fig.\ \ref{gfactor} is obtained by fitting
Eq.\ \ref{gfactormodel} to $\Delta X^+_S$. Using the measured values
of $t_h$ (0.86 meV) and $F_h$ (63.3 kV/cm) we find $g_{12}=1.65$.
The barrier contribution is positive, like the heavy hole g-factor
in bulk GaAs ($\sim 1.05$) \cite{KarasyukPRB94, SnellingPRB92}.
Because $g_T$ and $g_{12}$ have opposite sign, the splitting reaches
a minimum at $F_h$, where the amplitude in the barrier is at a
maximum. The g-factor at the minimum is $-0.44$.

Because the $\pm5/2$ triplet states recombine to the same bonding
orbital of the final hole (Fig.\ \ref{X+States}), the model predicts
the splitting of these lines as a function of electric field with no
additional fitting. As shown in Fig.\ \ref{BStates}c, the $\pm5/2$
triplet states have an initial state splitting of $g_e+2g_h$. The
transition g-factor is therefore given by
$g^B_T(F)=(g_e+2g_h)-(g_h+2abg_{12})=g_T-2abg_{12}$. This is shown
by the upper black line, which matches $\Delta X^+_{\pm5/2}$, the
observed splitting of the $\pm5/2$ triplets. The maximum splitting
corresponds to a g-factor of $-4.23$.

To apply the g-factor model to the bonding orbital of the neutral
exciton (lower branch Fig.\ \ref{Spectra}a) we use the measured
tunneling energy ($t_{X^0}$=0.58 meV) and anticrossing field
($F_{X^0}$=82.1 kV/cm). The lower red line in Fig.\ \ref{gfactor}
shows the fit to $\Delta X^0_B$, the Zeeman splitting of PL lines
from the bonding orbital. We find $g'_{12}=1.32$. The electron-hole
Coulomb interaction is responsible for the difference in tunneling
energy and anticrossing field from the bare hole case and is also
likely responsible for the difference between $g_{12}$ and
$g'_{12}$. The g-factor at the minimum is $-0.59$.

Using this value of $g'_{12}$, the model immediately explains the
increase in splitting for the antibonding orbital (upper branch
Fig.\ \ref{Spectra}a), which has a \textit{reduced} wavefunction
amplitude within the barrier. The g-factor for the antibonding
orbital is given by $g^A_T(F)=g_T-2abg'_{12}$, which
\textit{increases} in magnitude at the resonant field because $g_T$
is negative and $g'_{12}$ is positive. This is shown by the upper
red line in Fig.\ \ref{gfactor}, which matches $\Delta X^0_A$, the
Zeeman splitting for excitonic recombination from the antibonding
orbital. The splitting increases to a maximum (g-factor $-3.35$) at
the anticrossing point. The antibonding transitions for $X^+$ show
similar behavior, but are too weak to obtain full resonance curves.

The overall agreement between the model and experimental data is
quite good. There are some minor discrepancies, which highlight the
need for a detailed theory, possibly requiring inclusion of excited
states \cite{excitedstates}. However, the agreement of the data with
the resonance linewidths calculated using independently measured
values of $t_{X^0}$ and $t_h$ is strong confirmation that the
g-factor dependence does arise from the formation of bonding and
antibonding orbitals. For $X^+$ the g-factor resonance arises from
the wavefunction of the single hole in the final state, while for
$X^0$ the orbital wavefunction of the hole is influenced by the
additional electron.

We also studied samples in which electrons tunnel through the
barrier, with an anticrossing energy ($\sim 2.3$ meV) comparable to
that of the hole tunneling sample presented here ($\sim 1.7$ meV).
This requires a thicker barrier (10 nm) because of the smaller
electron effective mass. The electron wavefunction amplitudes in the
barrier should be at least as large as the hole-tunneling case.
However, the electron g-factor in bulk GaAs ($-0.44$)
\cite{KosakaElectLett01} is similar to the electron g-factor in InAs
quantum dots ($-0.6$) and according to our model the contribution
from the barrier should not significantly change the electron
g-factor. We see no electric field dependence of the g-factor in
these electron anticrossing samples. By adding aluminum to the
barrier, the resonant contribution to the electron g-factor could be
enhanced.

We have presented a resonant change in g-factor as a function of
electric field for the molecular spin states of QDMs with a thin
tunnel barrier. By studying single QDMs, we are able to identify the
individual molecular spin states and the different resonant behavior
of their g-factors. The results suggest that design of molecular
spin states and tunnel resonances may provide new opportunities for
combining optical and electrical control of confined spins.

\begin{acknowledgments}
We acknowledge financial support by NSA/ARO, CRDF, RFBR, RSSF, and
ONR. E.A.S., I.V.P., and M.F.D. are NRC/NRL Research Associates.
\end{acknowledgments}


\begin{thebibliography}{24}
\expandafter\ifx\csname
natexlab\endcsname\relax\def\natexlab#1{#1}\fi
\expandafter\ifx\csname bibnamefont\endcsname\relax
  \def\bibnamefont#1{#1}\fi
\expandafter\ifx\csname bibfnamefont\endcsname\relax
  \def\bibfnamefont#1{#1}\fi
\expandafter\ifx\csname citenamefont\endcsname\relax
  \def\citenamefont#1{#1}\fi
\expandafter\ifx\csname url\endcsname\relax
  \def\url#1{\texttt{#1}}\fi
\expandafter\ifx\csname urlprefix\endcsname\relax\def\urlprefix{URL
}\fi \providecommand{\bibinfo}[2]{#2}
\providecommand{\eprint}[2][]{\url{#2}}

\bibitem[{\citenamefont{Koppens et~al.}(2005)\citenamefont{Koppens, Folk,
  Elzerman, Hanson, van Beveren, Vink, Tranitz, Wegscheider, Kouwenhoven, and
  Vandersypen}}]{KoppensScience05}
\bibinfo{author}{\bibfnamefont{F.~H.~L.} \bibnamefont{Koppens, \textit{et~al.}}},
  \bibinfo{journal}{Science}
  \textbf{\bibinfo{volume}{309}}, \bibinfo{pages}{1346} (\bibinfo{year}{2005}).

\bibitem[{\citenamefont{Petta et~al.}(2005)\citenamefont{Petta, Johnson,
  Taylor, Laird, Yacoby, Lukin, Marcus, Hanson, and Gossard}}]{PettaScience05}
\bibinfo{author}{\bibfnamefont{J.~R.} \bibnamefont{Petta, \textit{et~al.}}},
  \bibinfo{journal}{Science}
  \textbf{\bibinfo{volume}{309}}, \bibinfo{pages}{2180} (\bibinfo{year}{2005}).

\bibitem[{\citenamefont{Loss and DiVincenzo}(1998)}]{LossPRA98}
\bibinfo{author}{\bibfnamefont{D.}~\bibnamefont{Loss}} \bibnamefont{and}
  \bibinfo{author}{\bibfnamefont{D.~P.} \bibnamefont{DiVincenzo}},
  \bibinfo{journal}{Phys. Rev. A} \textbf{\bibinfo{volume}{57}},
  \bibinfo{pages}{120} (\bibinfo{year}{1998}).

\bibitem[{\citenamefont{Vrijen et~al.}(2000)\citenamefont{Vrijen, Yablonovitch,
  Wang, Jiang, Balandin, Roychowdhury, Mor, and DiVincenzo}}]{VrijenPRA00}
\bibinfo{author}{\bibfnamefont{R.}~\bibnamefont{Vrijen, \textit{et~al.}}},
  \bibinfo{journal}{Phys. Rev. A} \textbf{\bibinfo{volume}{62}},
  \bibinfo{pages}{012306} (\bibinfo{year}{2000}).

\bibitem[{\citenamefont{Stinaff et~al.}(2006)\citenamefont{Stinaff, Scheibner,
  Bracker, Ponomarev, Korenev, Ware, Doty, Reinecke, and
  Gammon}}]{StinaffScience06}
\bibinfo{author}{\bibfnamefont{E.~A.} \bibnamefont{Stinaff, \textit{et~al.}}},
  \bibinfo{journal}{Science} \textbf{\bibinfo{volume}{311}},
  \bibinfo{pages}{636} (\bibinfo{year}{2006}).

\bibitem[{\citenamefont{Krenner et~al.}(2005)\citenamefont{Krenner, Sabathil,
  Clark, Kress, Schuh, Bichler, Abstreiter, and Finley}}]{KrennerPRL05}
\bibinfo{author}{\bibfnamefont{H.~J.} \bibnamefont{Krenner, \textit{et~al.}}},
  \bibinfo{journal}{Phys. Rev. Lett.}
  \textbf{\bibinfo{volume}{94}}, \bibinfo{pages}{057402}
  (\bibinfo{year}{2005}).

\bibitem[{\citenamefont{Krenner et~al.}(2006)\citenamefont{Krenner, Clark,
  Nakaoka, Bichler, Scheurer, Abstreiter, and Finley}}]{KrennerCondMat06}
\bibinfo{author}{\bibfnamefont{H.~J.} \bibnamefont{Krenner, \textit{et~al.}}},
  \bibinfo{journal}{cond-mat/0604659}
  (\bibinfo{year}{2006}).

\bibitem[{\citenamefont{Ortner et~al.}(2005)\citenamefont{Ortner, Bayer,
  Lyanda-Geller, Reinecke, Kress, Reithmaier, and Forchel}}]{OrtnerPRL05}
\bibinfo{author}{\bibfnamefont{G.}~\bibnamefont{Ortner, \textit{et~al.}}},
  \bibinfo{journal}{Phys. Rev. Lett.} \textbf{\bibinfo{volume}{94}},
  \bibinfo{pages}{157401} (\bibinfo{year}{2005}).

\bibitem[{\citenamefont{Scheibner et~al.}(2006)\citenamefont{Scheibner, Doty,
  Ponomarev, Bracker, Stinaff, Korenev, Reinecke, and Gammon}}]{Scheibner}
\bibinfo{author}{\bibfnamefont{M.}~\bibnamefont{Scheibner, \textit{et~al.}}},
  \bibinfo{journal}{cond-mat/0607241}  (\bibinfo{year}{2006}).

\bibitem[{\citenamefont{Salis et~al.}(2001)\citenamefont{Salis, Kato, Ensslin,
  Driscoll, Gossard, and Awschalom}}]{SalisNature01}
\bibinfo{author}{\bibfnamefont{G.}~\bibnamefont{Salis, \textit{et al.}}},
  \bibinfo{journal}{Nature}
  \textbf{\bibinfo{volume}{414}}, \bibinfo{pages}{619} (\bibinfo{year}{2001}).

\bibitem[{\citenamefont{Poggio et~al.}(2004)\citenamefont{Poggio, Steeves,
  Myers, Stern, Gossard, and Awschalom}}]{PoggioPRB04}
\bibinfo{author}{\bibfnamefont{M.}~\bibnamefont{Poggio, \textit{et~al.}}},
  \bibinfo{journal}{Phys. Rev. B}
  \textbf{\bibinfo{volume}{70}}, \bibinfo{pages}{121305(R)}
  (\bibinfo{year}{2004}).

\bibitem[{\citenamefont{Jiang and Yablonovitch}(2001)}]{JiangPRB01}
\bibinfo{author}{\bibfnamefont{H.~W.} \bibnamefont{Jiang}} \bibnamefont{and}
  \bibinfo{author}{\bibfnamefont{E.}~\bibnamefont{Yablonovitch}},
  \bibinfo{journal}{Phys. Rev. B} \textbf{\bibinfo{volume}{64}},
  \bibinfo{pages}{041307(R)} (\bibinfo{year}{2001}).

\bibitem[{\citenamefont{Lin et~al.}(2004)\citenamefont{Lin, Nitta, Koga, and
  Akazaki}}]{LinPhysicaE04}
\bibinfo{author}{\bibfnamefont{Y.}~\bibnamefont{Lin, \textit{et~al.}}},
  \bibinfo{journal}{Physica E} \textbf{\bibinfo{volume}{21}},
  \bibinfo{pages}{656} (\bibinfo{year}{2004}).

\bibitem[{gro()}]{growthdetails}
\bibinfo{note}{Dots are MBE grown on GaAs \cite{StinaffScience06} in a diode
  structure; an Al mask with 1 $\mu$m apertures isolates single QDMs.}

\bibitem[{num()}]{numericalvalues}
\bibinfo{note}{Numerical values from experimental spectra: $\Gamma_1=3.2$ meV,
  $\Gamma_2=13.1$ meV, $t_h=0.86$ meV, $t_{X^+}=1.24$ meV, $J^{eh}=0.116$ meV,
  $\tilde{d}=4.29$ nm.}

\bibitem[{Jhh()}]{Jhh}
\bibinfo{note}{We can not rule out a small direct exchange energy between holes
  in separate dots, $J^{hh}$. Because the anticrossing near 90 kV/cm is not
  fully observed, we can not separate $J^{hh}$ from the kinetic singlet-triplet
  splitting due to tunneling. Since the kinetic splitting is dominant in these
  samples we have, for simplicity, taken $J^{hh}=0$.}

\bibitem[{STm()}]{STmixing}
\bibinfo{note}{The $^{\uparrow,0}_{\Downarrow, \Uparrow}X^+_{\pm1/2}$ states
  mix slightly with $^{\uparrow,0}_{\Downarrow, \Uparrow}X^+_S$ through
  electron-hole exchange, $J^{eh}$ \cite{Scheibner}.}

\bibitem[{\citenamefont{Fazekas}(1999)}]{Fazekas99}
\bibinfo{author}{\bibfnamefont{P.}~\bibnamefont{Fazekas}},
  \emph{\bibinfo{title}{{Lecture Notes on Electron Correlation and Magnetism}}}
  (\bibinfo{publisher}{World Scientific, Singapore}, \bibinfo{year}{1999}).

\bibitem[{\citenamefont{Bayer et~al.}(2000)\citenamefont{Bayer, Stern, Kuther,
  and Forchel}}]{BayerPRB00}
\bibinfo{author}{\bibfnamefont{M.}~\bibnamefont{Bayer, \textit{et~al.}}},
  \bibinfo{journal}{Phys. Rev. B} \textbf{\bibinfo{volume}{61}},
  \bibinfo{pages}{7273} (\bibinfo{year}{2000}).

\bibitem[{gfa()}]{gfactors}
\bibinfo{note}{Our values of $g_e=-0.6$ and $g_h=-1.6$ are typical for InAs
  QDs. For bulk GaAs, $g_h \sim 1.05$ for heavy holes and $g_e=-0.44$
  \cite{KarasyukPRB94}. Zeeman splittings for both electrons and holes are
  given by $g \mu_B B$.}

\bibitem[{\citenamefont{Karasyuk et~al.}(1994)\citenamefont{Karasyuk, Beckett,
  Nissen, Villemaire, Steiner, and Thewalt}}]{KarasyukPRB94}
\bibinfo{author}{\bibfnamefont{V.~A.} \bibnamefont{Karasyuk, \textit{et~al.}}},
  \bibinfo{journal}{Phys. Rev. B}
  \textbf{\bibinfo{volume}{49}}, \bibinfo{pages}{16381} (\bibinfo{year}{1994}).

\bibitem[{\citenamefont{Snelling et~al.}(1992)\citenamefont{Snelling,
  Blackwood, McDonagh, Harley, and Foxon}}]{SnellingPRB92}
\bibinfo{author}{\bibfnamefont{M.~J.} \bibnamefont{Snelling, \textit{et~al.}}},
  \bibinfo{journal}{Phys. Rev. B}
  \textbf{\bibinfo{volume}{45}}, \bibinfo{pages}{3922(R)}
  (\bibinfo{year}{1992}).

\bibitem[{exc()}]{excitedstates}
\bibinfo{note}{There is evidence that the first excited hole state is $\sim$7
  meV higher in energy. ($\Delta F\sim 16$ kV/cm)}.

\bibitem[{\citenamefont{Kosaka et~al.}(2001)\citenamefont{Kosaka, Kiselev,
  Baron, Kim, and Yablonovitch}}]{KosakaElectLett01}
\bibinfo{author}{\bibfnamefont{H.}~\bibnamefont{Kosaka, \textit{et~al.}}},
  \bibinfo{journal}{Elect. Lett.} \textbf{\bibinfo{volume}{37}},
  \bibinfo{pages}{464} (\bibinfo{year}{2001}).

\end{thebibliography}
\end{document}